\documentclass{JHEP3}
\usepackage{graphicx}
\usepackage{amsmath}
\usepackage{psfrag}
\usepackage[sort&compress,numbers]{natbib}

\def\eps{\epsilon}
\def\CA{C_A}
\def\CF{C_F}

\def\NF{N_F}
\def\NFV{N_{F,V}}
\def\NN{\left(\frac{N^2-4}{N}\right)}

\def\e{\epsilon}

\preprint{
ZU--TH 15/10\\  IPPP/10/88\\ SI-HEP-2010-16\\ BI-TP 2010/38}

\title{The quark and gluon form factors to three loops in QCD through to {\cal O}$(\eps^2)$}

\author{T.\  Gehrmann$^a$, E.W.N. Glover$^b$, T.\ Huber$^{c}$,
   N.\ Ikizlerli$^b$, C.\ Studerus$^{a,d}$
	\\
$^a$ Institut f\"ur Theoretische Physik, Universit\"at Z\"urich, Winterthurerstrasse 190, CH-8057 Z\"urich, 
Switzerland\\
	$^b$Institute for Particle Physics Phenomenology, University of Durham,
South Road,\\ Durham DH1 3LE, England\\
$^c$ Fachbereich 7, Universit\"at Siegen, Walter-Flex-Strasse 3, D-57068 Siegen, Germany\\
$^d$ Faculty of Physics, University of Bielefeld, D-33501 Bielefeld, Germany}

\abstract{We give explicit formulae for the ${\cal O}(\eps)$ and ${\cal O}(\eps^2)$ contributions to the unrenormalised three loop QCD corrections to quark and gluon form factors.  These contributions have at most transcendentality weight eight. 
The ${\cal O}(\e)$ terms of 
the three-loop form factors are required for the extraction of the four-loop 
quark and gluon collinear anomalous dimensions.
The ${\cal O}(\e^2)$ terms represent an irreducible contribution to the finite part of the form factors at four-loops.
For the sake of completeness, we also give the contributions to the one and two loop
form factors to the same transcendentality weight eight.
 }

\begin{document}

The form factors are fundamental ingredients for many precision  calculations in
QCD. These basic building blocks describe the coupling of an external,
colour-neutral off-shell particle to a pair of partons:  the quark form factor is
the coupling of a virtual photon to a quark-antiquark pair, while the gluon form
factor is the coupling of a Higgs boson to a pair of gluons through an effective
Lagrangian.  

The form factors are phenomenologically important and appear directly as virtual
higher-order corrections in coefficient functions for the inclusive Drell-Yan 
process~\cite{dy1l,dy2l,hk1}  and the inclusive Higgs production cross
section~\cite{hk1,higgs1l,higgsfull,higgs2l}.  The form factors also display a
non-trivial infrared pole structure which is determined by the infrared
factorisation formula. This implies that their infrared pole coefficients can be
used to extract fundamental constants such as the cusp anomalous dimensions which
control the structure of soft divergences and the collinear quark and gluon
anomalous dimensions.  In fact,  the cusp anomalous dimensions were 
first obtained to three loops from 
the asymptotic behaviour of splitting 
functions~\cite{Moch:2004pa,Vogt:2004mw}.  However, it was   
 the calculation~\cite{MMV1,MMV2} of the 
pole terms of the three-loop form factors 
(and finite plus subleading terms in the 
two-loop and one-loop form factors~\cite{vn,harlander,GHM}), which 
led to the derivation of the three-loop collinear 
anomalous dimensions~\cite{MMV1,Becher:2006mr,Becher:2009qa}.

The infrared factorisation formula for a given form factor (or more generally
for a given multi-leg amplitude) at a certain number of loops involves infrared
singularity operators acting on the form factor evaluated with a lower number of
loops. These infrared singularity operators contain explicit infrared poles
$1/\eps^2$ and $1/\eps$. Therefore, the computation of the finite contribution to any $n$-loop form factor relies on contributions from $(n-m)$-loops evaluated to ${\cal O}(\eps^{2m})$.  

\begin{figure}[t]
\psfrag{b41}{$B_{4,1}~[=A_4]$}
\psfrag{b51}{$B_{5,1}~[=A_{5,3}]$}
\psfrag{b52}{$B_{5,2}~[=A_{5,4}]$}
\psfrag{b61}{$B_{6,1}~[=A_{6,6}]$}
\psfrag{b62}{$B_{6,2}~[=A_{6,4}]$}
\psfrag{b81}{$B_{8,1}$}
\psfrag{c61}{$C_{6,1}~[=A_{6,5}]$}
\psfrag{c81}{$C_{8,1}$}
\psfrag{a51}{$A_{5,1}$}
\psfrag{a52}{$A_{5,2}$}
\psfrag{a61}{$A_{6,1}$}
\psfrag{a62}{$A_{6,2}$}
\psfrag{a63}{$A_{6,3}$}
\psfrag{a71}{$A_{7,1}$}
\psfrag{a72}{$A_{7,2}$}
\psfrag{a73}{$A_{7,3}$}
\psfrag{a74}{$A_{7,4}$}
\psfrag{a75}{$A_{7,5}$}
\psfrag{a81}{$A_{8,1}$}
\psfrag{a91}{$A_{9,1}$}
\psfrag{a92}{$A_{9,2}$}
\psfrag{a94}{$A_{9,4}$}
	\begin{center}
		\includegraphics[height=2cm]{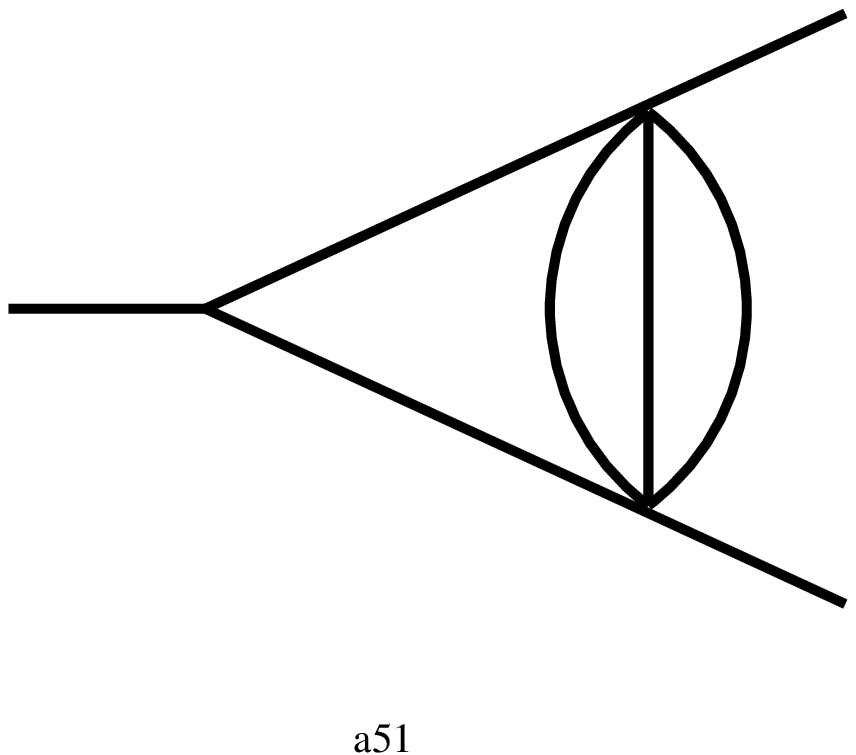}
		\hspace{1cm}
		\includegraphics[height=2cm]{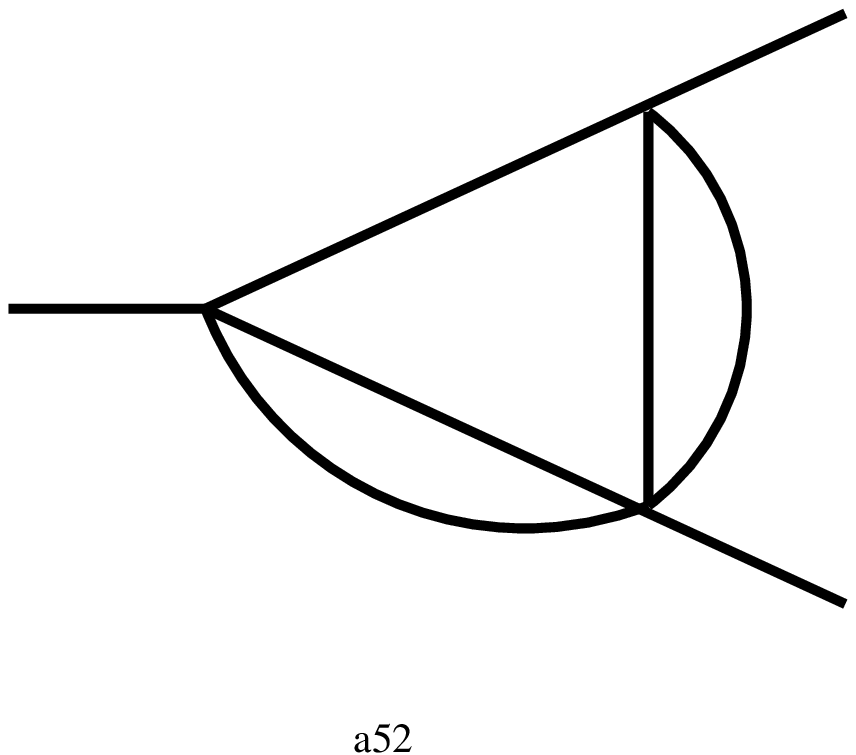}
		\hspace{1cm}
		\includegraphics[height=2cm]{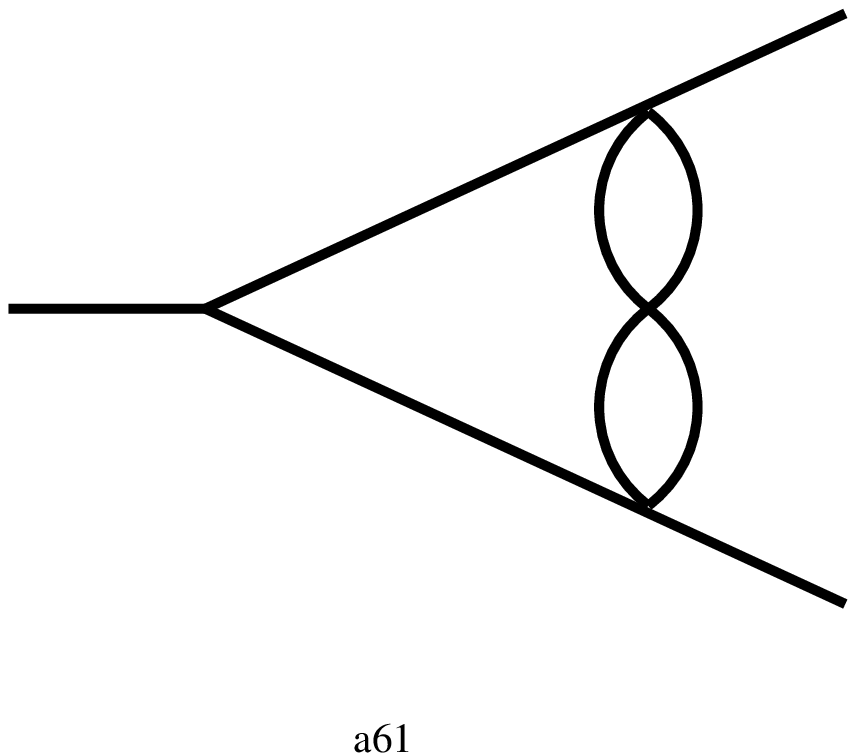}
		\hspace{1cm}
		\includegraphics[height=2cm]{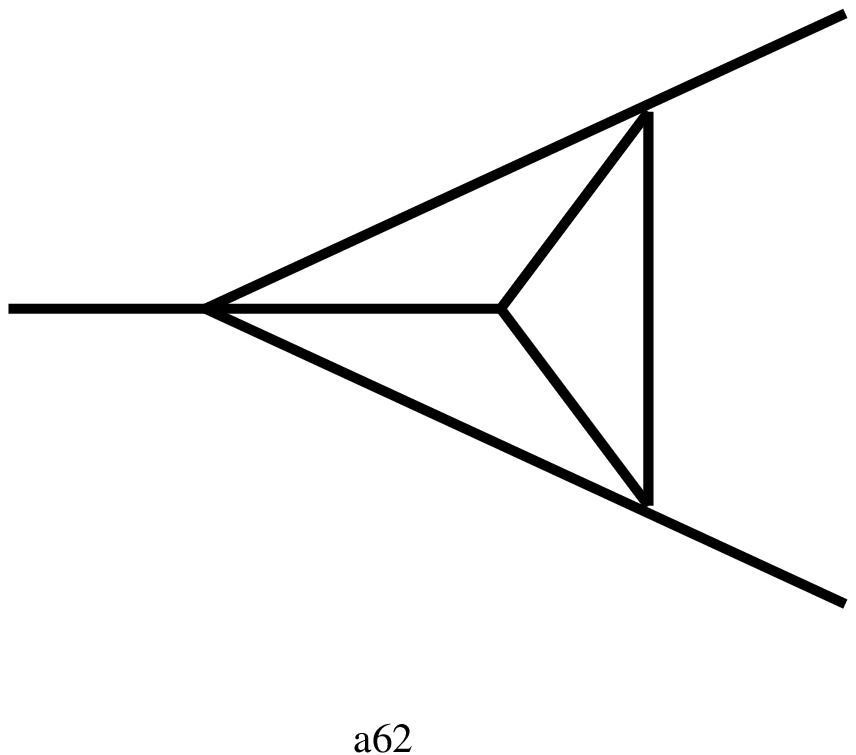}
	\end{center}	 
	\vspace{0.5cm}
	\begin{center}
		\hspace{1cm}
		\includegraphics[height=2cm]{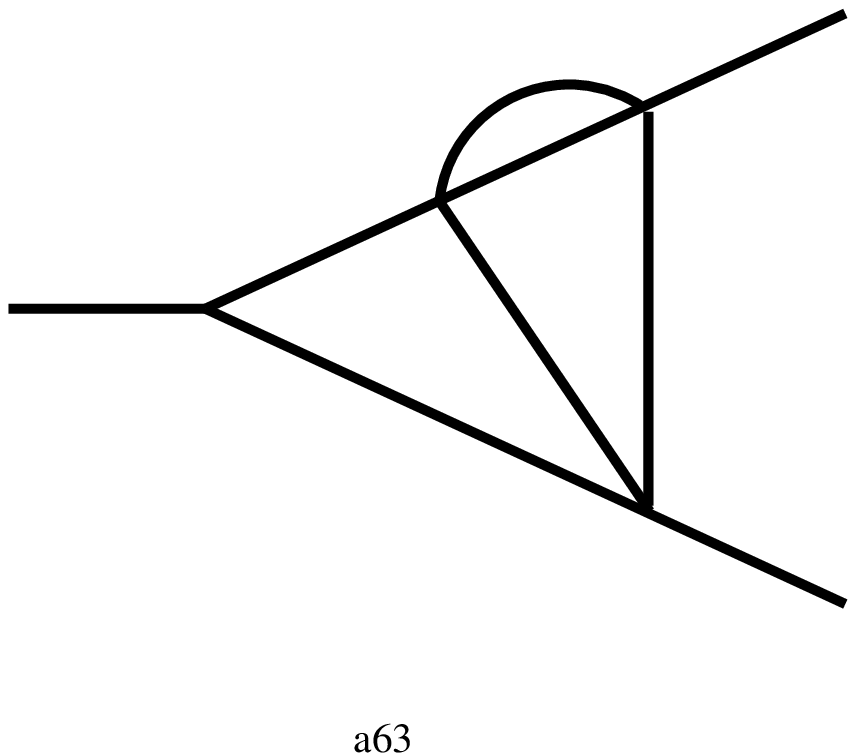}
		\hspace{1cm}
		\includegraphics[height=2cm]{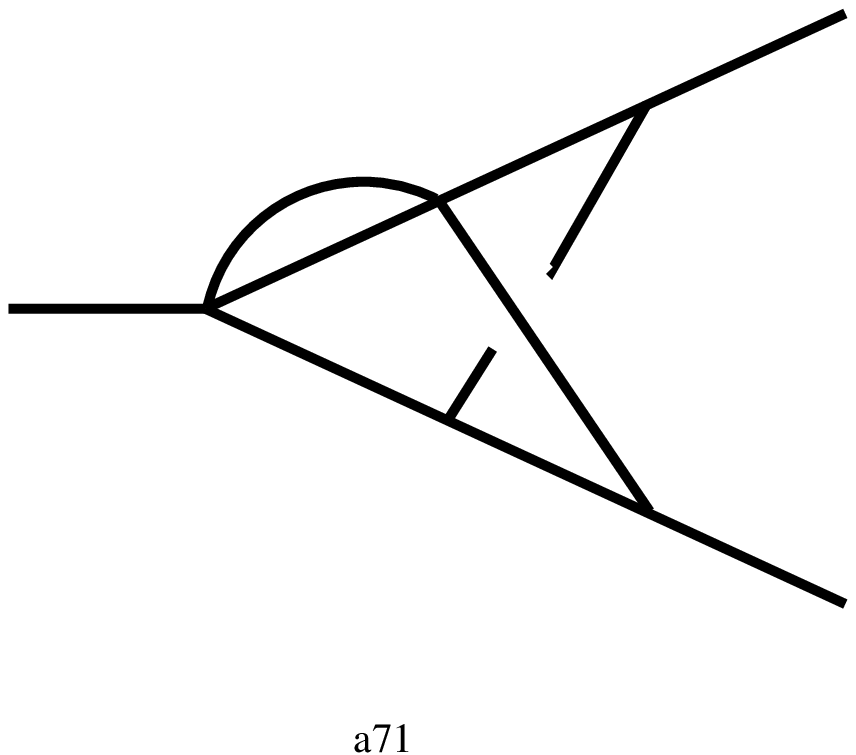}
		\hspace{1cm}
		\includegraphics[height=2cm]{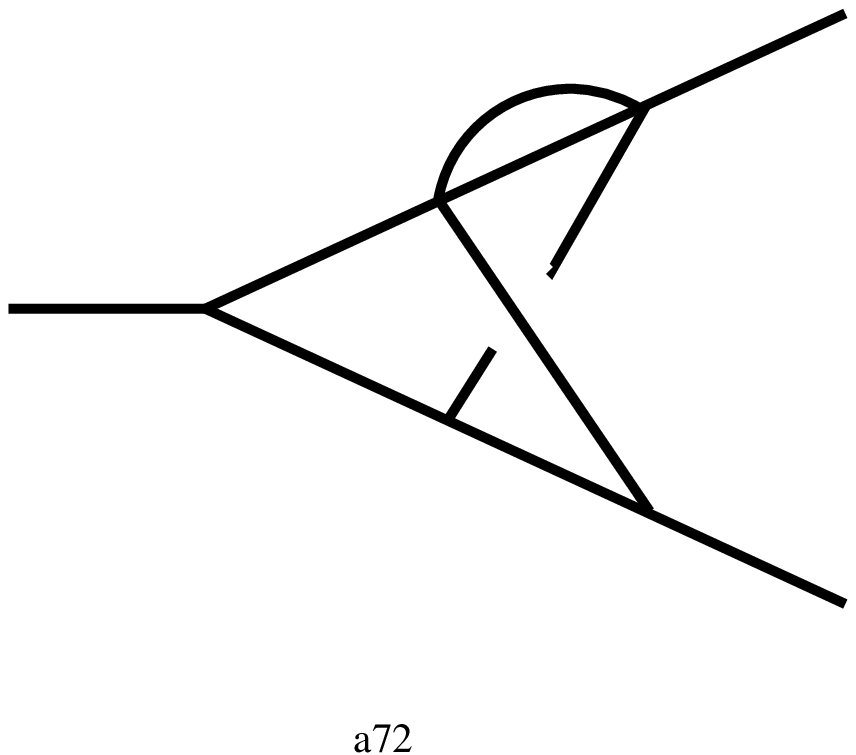}
		\hspace{1cm}
		\includegraphics[height=2cm]{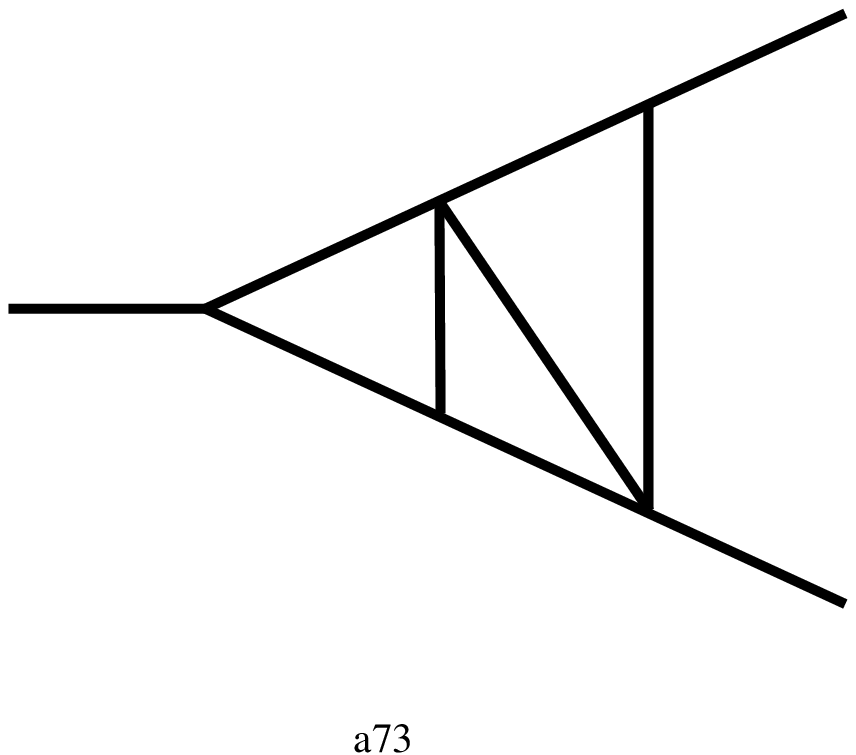}
	\end{center}	 
	\vspace{0.5cm}
	\begin{center}
		\includegraphics[height=2cm]{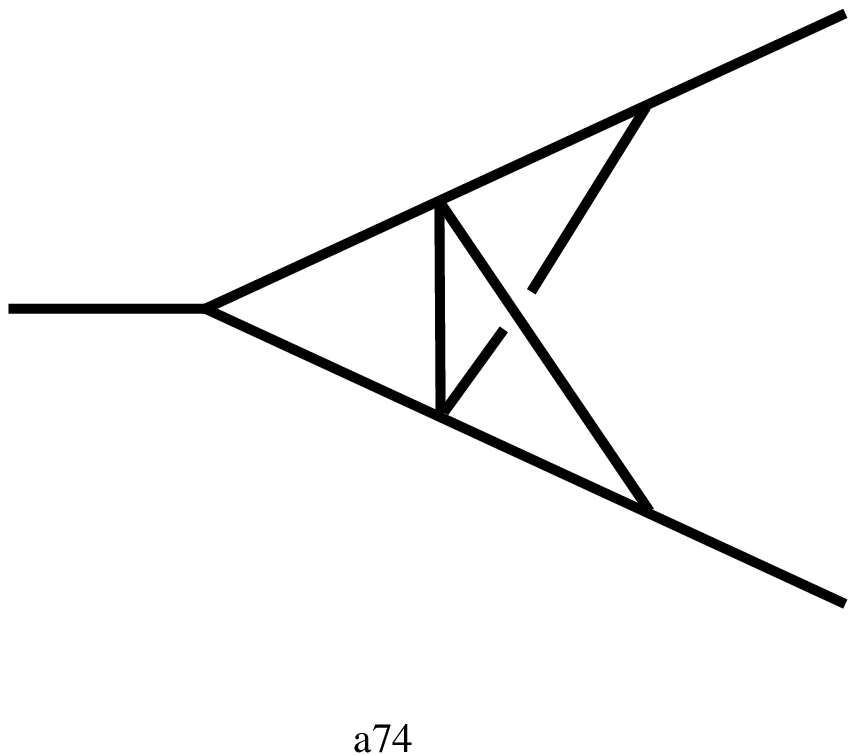}
		\hspace{1cm}
		\includegraphics[height=2cm]{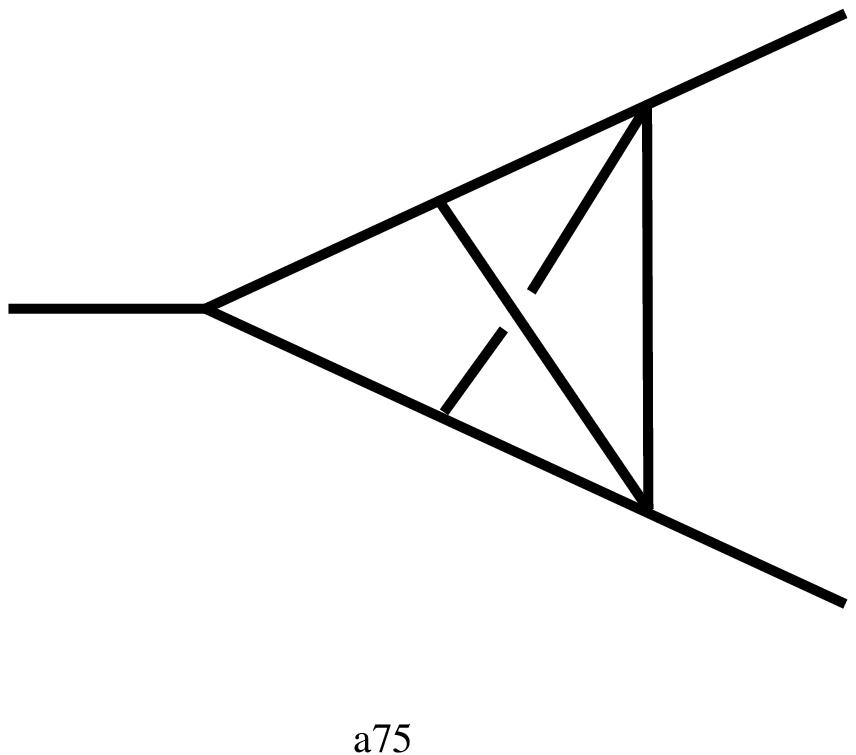}
		\hspace{1cm}
		\includegraphics[height=2cm]{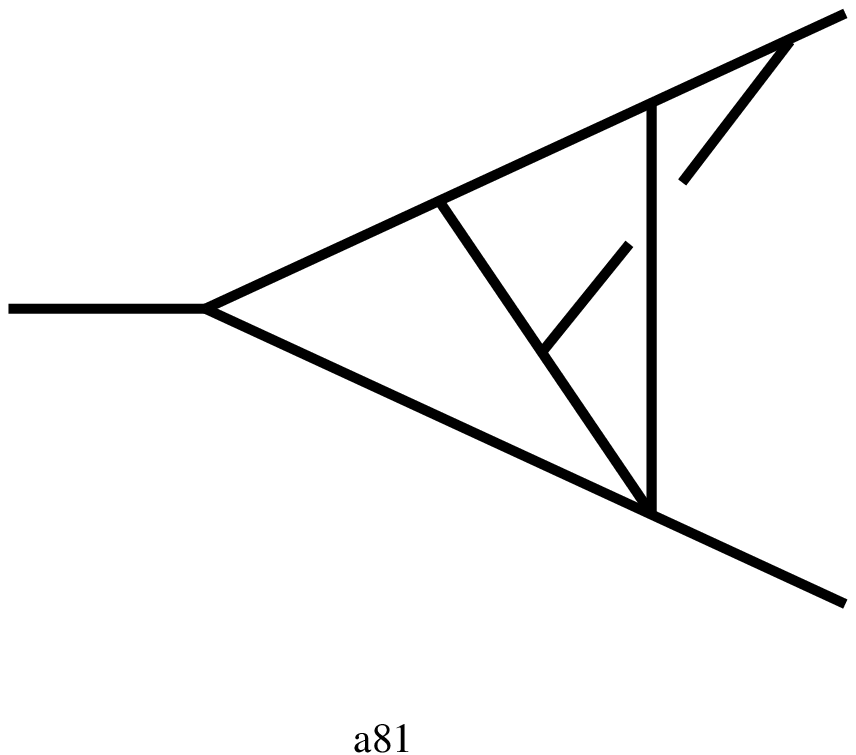}
		\hspace{1cm}
		\includegraphics[height=2cm]{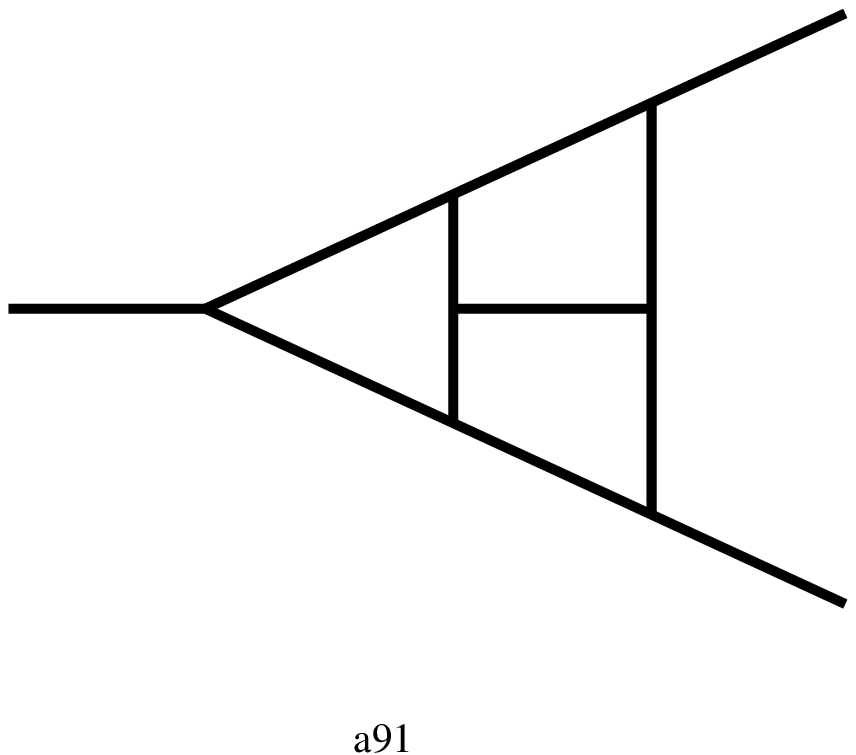}
	\end{center}	 
	\vspace{0.5cm}
	\begin{center}
		\includegraphics[height=2cm]{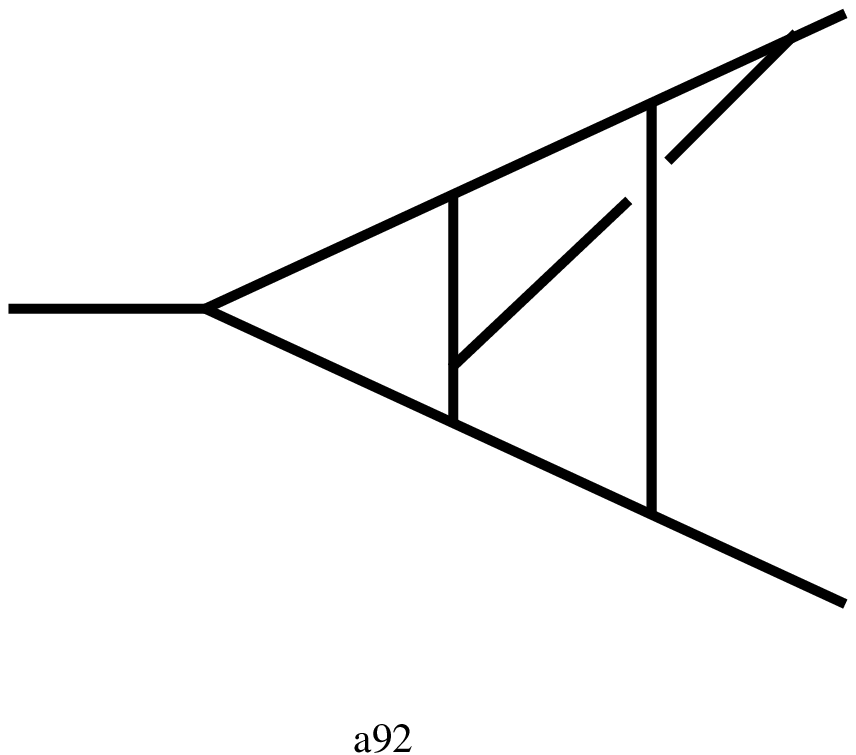}
		\hspace{1cm}
		\includegraphics[height=2cm]{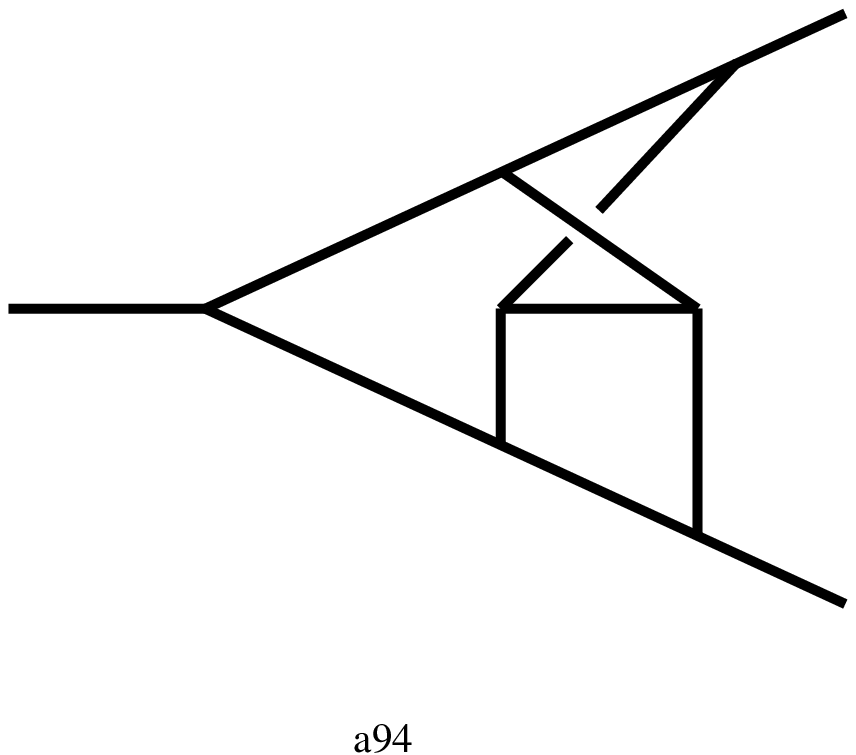}
	\end{center}	 
	\vspace{0.5cm}
	\begin{center}
		\includegraphics[height=1.8cm]{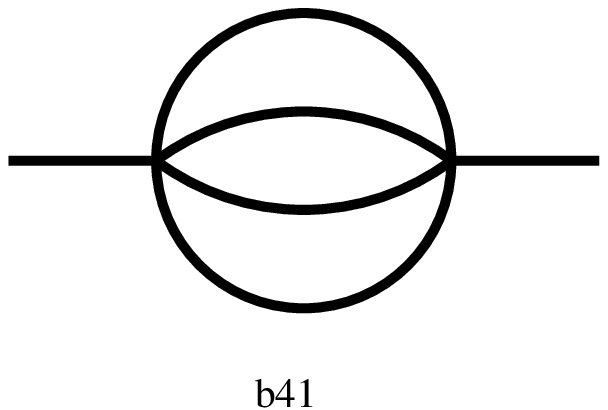}
		\hspace{1cm}
		\includegraphics[height=1.8cm]{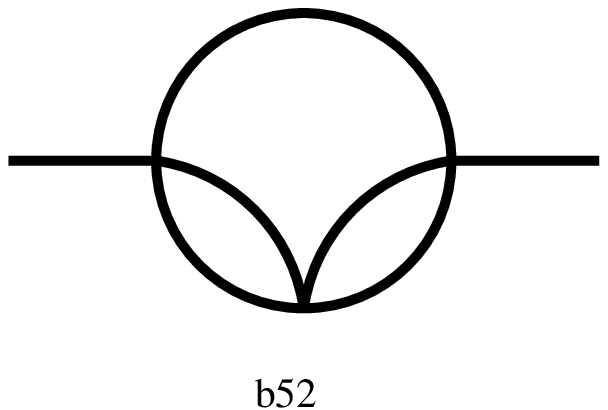}
		\hspace{1cm}
		\includegraphics[height=1.8cm]{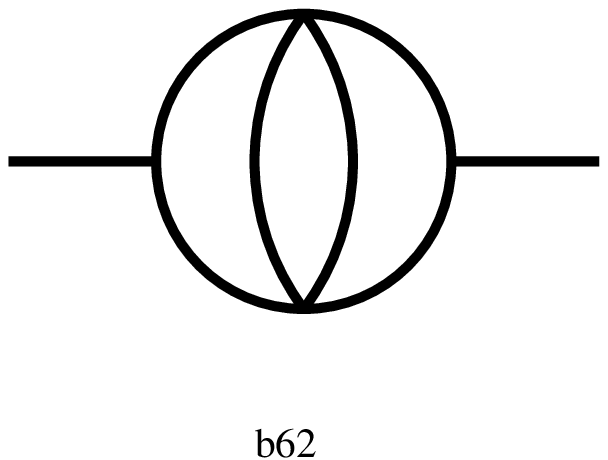}
		\hspace{1cm}
		\includegraphics[height=1.8cm]{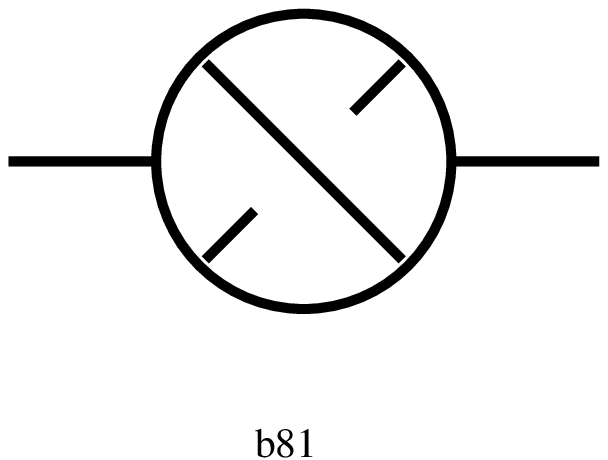}
	\end{center}	 
	\vspace{0.5cm}
	\begin{center}
		\includegraphics[height=1.2cm]{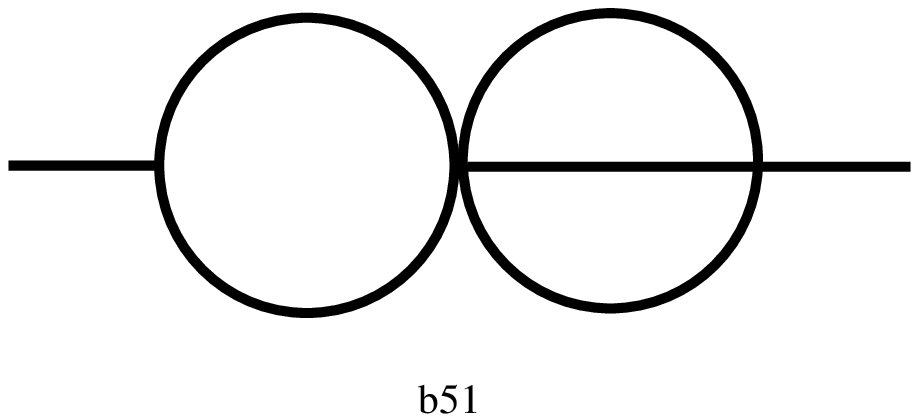}
		\hspace{1cm}
		\includegraphics[height=1.2cm]{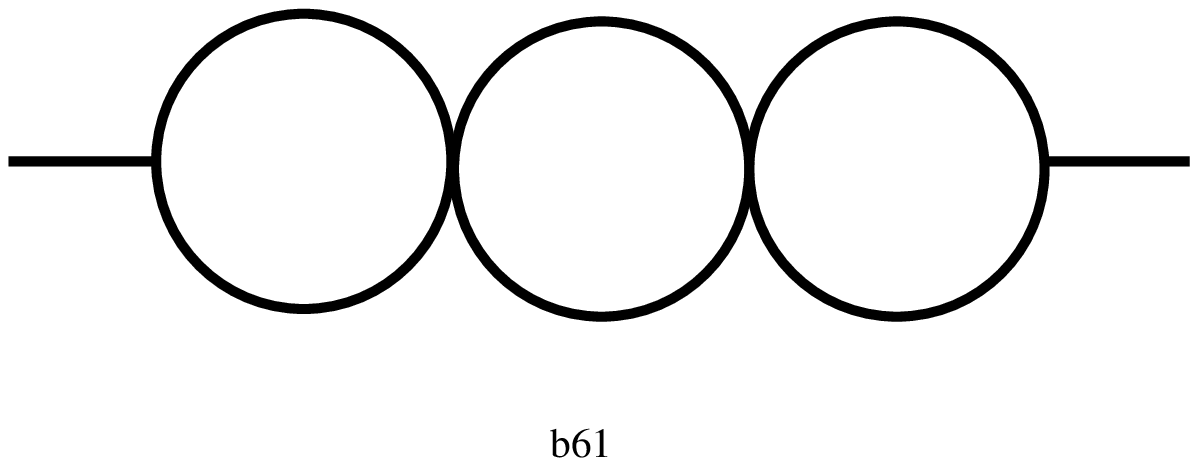}
		\hspace{1cm}
		\includegraphics[height=1.2cm]{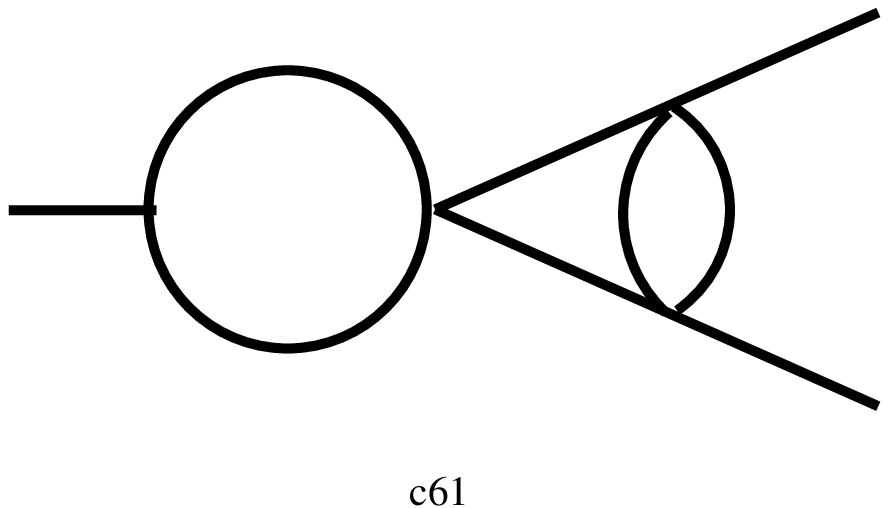}
		\hspace{1cm}
		\includegraphics[height=1.2cm]{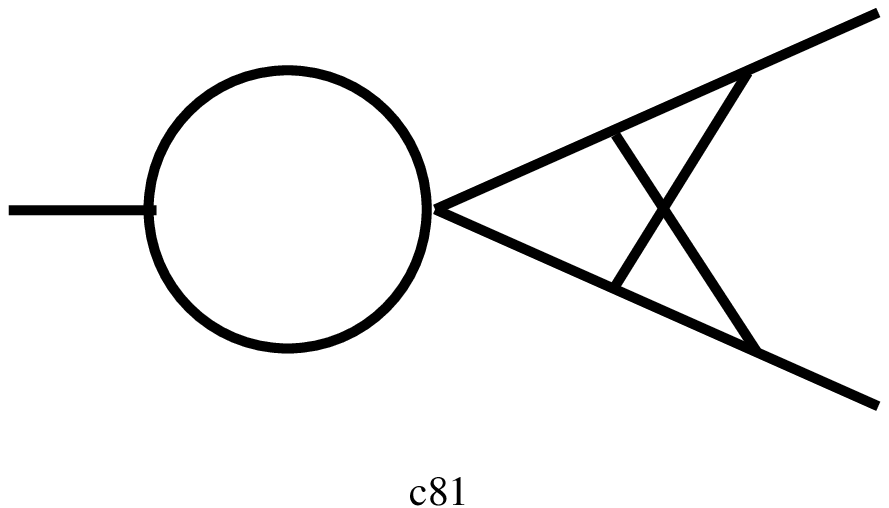}
	\end{center}	 
\caption{Master integrals for the three-loop form factors.  Labels in brackets indicate the naming convention of Ref.~\cite{leesmirnov}.}
\label{fig:MI}
\end{figure}

At present, the state of the art is at the three-loop level for the massless quark and gluon form factors. There are 22 master integrals  shown in Fig.~\ref{fig:MI}, of which 14 are
genuine three-loop vertex  functions ($A_{t,i}$-type), 4 are three-loop propagator
integrals ($B_{t,i}$-type) and 4 are products of one-loop and two-loop integrals
($C_{t,i}$-type).  In this notation, the index $t$ denotes the number of propagators,
and $i$ is  simply enumerating the topologically different integrals with the same
number of propagators.
 Expressions for the form factors in terms of the 22 independent master integrals, and valid for any value of the dimension $D$,  are given in Ref.~\cite{GGHIS}.   The $B_{t,i}$-type integrals were computed to finite order
in~\cite{chet1,mincer} and supplemented by the higher order  terms
in~\cite{bekavac}. Explicit expansions 
of the $A_{t,i}$-type integrals were obtained in Refs.~\cite{masterA,masterB,masterC,masterD} using Mellin-Barnes techniques. They enabled the evaluation of the three-loop form factors up to and including the finite contributions~\cite{BCSSS,masterD,GGHIS}.   The deepest pole contribution is of ${\cal O}(1/\eps^{6})$.   Correspondingly, the finite terms are of at most transcendentality weight six,  that is terms such as $\pi^6$ $(\zeta_2^3)$ or $\zeta_3^2$.  

More recently~\cite{leesmirnov}, 20 of the three-loop master  integrals have been
re-evaluated up to transcendentality weight eight
using dimensional recurrence relations~\cite{Tarasov:1996br,Lee:2009dh} and
analytic properties of Feynman integrals (the DRA method~\cite{leesmirnov2}).
Expressions for the two remaining integrals, $B_{8,1}$ and $C_{8,1}$,
can be obtained from Refs.~\cite{leesmirnov2} and \cite{GHM} respectively.
Once the same normalisation and basis set of multiple zeta values is used, 
Ref.~\cite{leesmirnov} confirms the earlier result of Ref.~\cite{HypExp2} for $A_{6,2}$.
On the other hand, we confirm a certain subset of master integrals ($B_{6,2}$, $B_{8,1}$, $A_{7,3}$, $A_{7,5}$,
$A_{8,1}$, $A_{9,1}$, $A_{9,2}$, $A_{9,4}$) from~\cite{leesmirnov,leesmirnov2}
up to coefficients corresponding to weight eight
numerically to a precision of one per-mille or better using {\tt MB.m}~\cite{Czakon:2005rk} and {\tt FIESTA}~\cite{Smirnov:2008py,Smirnov:2009pb}.
All other of the 22 master integrals we even confirm analytically through to weight eight
by expanding the closed form in terms of hypergeometric functions
given in~\cite{masterA,masterB} using the {\tt HypExp} package~\cite{HypExp}.
 
All master integrals are therefore known up to transcendentality weight eight i.e.\ terms including $\pi^8$ $(\zeta_2^4)$, $\zeta_2\zeta_3^2$, $\zeta_3\zeta_5$ as well as the multiple zeta value $\zeta_{5,3}$ (or equivalently $\zeta_{-6,-2}$). 
The multiple zeta values are defined by (see e.g.~\cite{datamine} and references therein)
\begin{equation}\label{MZVdef}
\zeta(m_1,\dots,m_k)=\sum\limits_{i_1=1}^\infty\sum\limits_{i_2=1}^{i_1-1}
\dots\sum\limits_{i_k=1}^{i_{k-1}-1}\prod\limits_{j=1}^k\frac{\mbox{sgn}(m_j)^{i_j}}{i_j^{|m_j|}}\,.
\end{equation}
Specifically,  $\zeta_{-6,-2}$ is related to $\zeta_{5,3}$ by~\cite{datamine,HPL}
\begin{equation}
\zeta_{-6,-2} = \frac{9}{20}\zeta_{5,3}-\frac{3}{2}\zeta_{5}\zeta_{3}+\frac{781}{4032000}\pi^8.
\end{equation}
The numerical values of the transcendental constants up to weight eight are:
\begin{eqnarray*}
&& \zeta_3 = 1.2020569031595942854\ldots\,, \qquad 
\zeta_5= 1.0369277551433699263\ldots\, ,\\
&& \zeta_7 = 1.0083492773819228268\ldots\,,\qquad 
\zeta_{5,3} = 0.037707672984847544011\ldots\, .
\end{eqnarray*}                      
The new results on the higher order terms in the master integrals  enable the computation of the three-loop form factors through to ${\cal O}(\eps^2)$ which is an intrinsic component for the 
four-loop evaluation of the form factors.   This is the topic of this Letter and we give explicit formulae for the ${\cal O}(\eps)$ and ${\cal O}(\eps^2)$ contributions to the unrenormalised three loop form factors.

The form factors are the basic vertex functions of an external off-shell current 
(with virtuality $q^2 = s_{12}$) coupling to a pair of partons with on-shell
momenta $p_1$ and $p_2$. One distinguishes time-like ($s_{12}>0$, i.e.\ with
partons both either in the initial or in the final state) and  space-like
($s_{12}<0$, i.e.\ with one parton in the initial and one in the final state)
configurations.  The form factors are described in terms of scalar functions by
contracting the respective vertex functions  (evaluated in dimensional
regularization with $D=4-2\e$ dimensions) with  projectors.  For massless
partons, the full vertex function is described with only a single form factor. 

The quark form factor is obtained from the photon-quark-antiquark vertex $\Gamma^\mu_{q\bar q}$ by
\begin{equation}
{\cal F}^q  = -\frac{1}{4(1-\e)q^2}\, {\mathrm Tr} \left( p_2 \!\!\!\! / \, \Gamma^\mu_{q\bar q} p_1 \!\!\!\! / \, \gamma_\mu\right)\, , \label{eq:projq}
\end{equation}
while the gluon form factor relates to the effective Higgs-gluon-gluon vertex $\Gamma^{\mu \nu}_{gg}$ as
\begin{equation}
{\cal F}^g = \frac{p_1\cdot p_2 \, g_{\mu\nu} - p_{1,\mu} p_{2,\nu} - p_{1,\nu} p_{2,\mu} }{2 (1-\e)} \, 
\Gamma_{gg}^{\mu \nu}\, . \label{eq:projg}
\end{equation}
The form factors are expanded in perturbative QCD in powers of the coupling constant, with each power 
corresponding to a virtual loop. We denote the unrenormalized form factors by ${\cal F}^a$ and the renormalized 
form factors by $F^a$ with $a=q,g$. 

At tree level, the Higgs boson does not couple either to the gluon or to massless quarks. In higher orders in perturbation theory, heavy quark loops introduce
a coupling between the Higgs boson and gluons. In the limit of infinitely 
massive quarks, these loops give rise to an effective Lagrangian~\cite{hgg} 
mediating the 
coupling between the scalar Higgs field and the gluon field strength tensor:
\begin{equation}
{\cal L}_{{\rm int}} = -\frac{\lambda}{4} H F_a^{\mu\nu} F_{a,\mu\nu}\ .
\label{eq:lagr}
\end{equation}
The coupling $\lambda$ has inverse mass dimension. It can be computed 
by matching~\cite{kniehl,kniehl2} 
the effective theory to the full standard 
model cross sections~\cite{higgsfull}.

Direct evaluation of the Feynman diagrams at the appropriate loop order yields the bare (unrenormalised) form factors,
\begin{eqnarray}
{\cal F}_b^q (\alpha_s^b, s_{12}) &=& 1 + \sum_{n=1}^{\infty} \left( \frac{\alpha_s^b}{4\pi}\right)^n \left(\frac{-s_{12}}{\mu_0^2}\right)^{-n\eps} S_{\eps}^n \, {\cal F}_n^q,\\
{\cal F}_b^g (\alpha_s^b, s_{12}) &=& \lambda^b\left(1 + \sum_{n=1}^{\infty} \left( \frac{\alpha_s^b}{4\pi}\right)^n \left(\frac{-s_{12}}{\mu_0^2}\right)^{-n\eps} 
S_{\eps}^n \,{\cal F}_n^g\right),
\end{eqnarray}
where $\mu_0^2$ is the mass parameter introduced in dimensional regularisation to maintain a dimensionless coupling in the bare Lagrangian density and where
\begin{equation}
S_{\eps} = e^{-\eps \gamma} (4\pi)^{\eps}, \qquad \qquad {\rm with~the~Euler~constant~} \gamma =  0.5772\ldots
\end{equation}

The one-loop and two-loop form factors were computed in many places in the
literature~\cite{vn,harlander,GHM,MMV1,MMV2}.  All-order expressions  in terms of
one-loop and two-loop master integrals are given in~\cite{GHM}. Explicit expressions for
the one- and two-loop form factors through to {\cal O}($\eps^5)$ and {\cal O}($\eps^3)$
respectively are given already in ~\cite{GGHIS}. To determine the finite piece at 
the four-loop level, these form factors are needed to one higher power in $\eps$, and for the sake of completeness, we quote them here.  At one-loop,
{\allowdisplaybreaks
\begin{eqnarray}
{\cal F}_1^q &=& {\cal F}_1^q|_{\frac{1}{\eps^2}} + \ldots + {\cal F}_1^q|_{\eps^5}\nonumber \\
&+& \CF \Biggl [ \eps^6\biggl(-512
+\frac{381\zeta_7}{7}
+\frac{496\zeta_5}{5}
+\frac{448\zeta_3}{3}
-\frac{434\zeta_3\zeta_5}{15}
-\frac{196\zeta_3^2}{9}
+64\zeta_2
\nonumber \\ &&\hspace{1cm}
-\frac{93\zeta_2\zeta_5}{10}
-\frac{56\zeta_2\zeta_3}{3}
+\frac{49\zeta_2\zeta_3^2}{18}
+\frac{188\zeta_2^2}{5}
-\frac{329\zeta_2^2\zeta_3}{40}
+\frac{949\zeta_2^3}{70}
+\frac{55779\zeta_2^4}{11200}
\biggr)\Biggr ]  \nonumber \\
{\cal F}_1^g &=& {\cal F}_1^g|_{\frac{1}{\eps^2}} + \ldots + {\cal F}_1^g|_{\eps^5}\nonumber \\ 
&+& \CA \Biggl [ \eps^6\biggl(-126
+\frac{62\zeta_5}{5}
+\frac{98\zeta_3}{3}
-\frac{434\zeta_3\zeta_5}{15}
+15\zeta_2
-\frac{7\zeta_2\zeta_3}{3}
+\frac{49\zeta_2\zeta_3^2}{18}
\nonumber \\ &&\hspace{1cm}
+\frac{141\zeta_2^2}{20}
+\frac{55779\zeta_2^4}{11200}
\biggr)\Biggr ]  
\end{eqnarray}
}
and at two-loops
{\allowdisplaybreaks
\begin{eqnarray}
{\cal F}_2^q &=& {\cal F}_2^q|_{\frac{1}{\eps^4}} + \ldots + {\cal F}_2^q|_{\eps^3}\nonumber \\
&+& \CF^2 \Biggl [ \eps^4\biggl(+\frac{637631}{128}
-528\zeta_{5,3}
+\frac{27204\zeta_7}{7}
-\frac{34001\zeta_5}{10}
-\frac{481913\zeta_3}{24}
+\frac{33248\zeta_3\zeta_5}{15}
\nonumber \\ &&\hspace{1cm}
+\frac{36359\zeta_3^2}{9}
+\frac{95559\zeta_2}{32}
-198\zeta_2\zeta_5
+\frac{2257\zeta_2\zeta_3}{2}
-\frac{4576\zeta_2\zeta_3^2}{9}
-\frac{248023\zeta_2^2}{80}
\nonumber \\ &&\hspace{1cm}
+\frac{5109\zeta_2^2\zeta_3}{5}
+\frac{55623\zeta_2^3}{140}
+\frac{653901\zeta_2^4}{700}
\biggr)\Biggr ] \nonumber \\ 
&+& \CF \CA \Biggl [ \eps^4\biggl(-\frac{11630115085}{839808}
+264\zeta_{5,3}
-\frac{11980\zeta_7}{21}
+\frac{1214029\zeta_5}{270}
+\frac{84520897\zeta_3}{5832}
\nonumber \\ &&\hspace{1cm}
-\frac{8266\zeta_3\zeta_5}{5}
-\frac{229042\zeta_3^2}{81}
-\frac{58499773\zeta_2}{23328}
-\frac{829\zeta_2\zeta_5}{15}
-\frac{94931\zeta_2\zeta_3}{162}
+\frac{3029\zeta_2\zeta_3^2}{9}
\nonumber \\ &&\hspace{1cm}
+\frac{14915741\zeta_2^2}{6480}
-\frac{66379\zeta_2^2\zeta_3}{90}
+\frac{4843\zeta_2^3}{30}
-\frac{75242\zeta_2^4}{175}
\biggr)\Biggr ] \nonumber \\ 
&+& \CF\NF \Biggl [ \eps^4\biggl(+\frac{996726245}{419904}
-\frac{2186\zeta_7}{21}
-\frac{42713\zeta_5}{135}
-\frac{1951625\zeta_3}{2916}
+\frac{4732\zeta_3^2}{81}
+\frac{2877653\zeta_2}{11664}
\nonumber \\ &&\hspace{1cm}
-\frac{242\zeta_2\zeta_5}{15}
-\frac{4589\zeta_2\zeta_3}{81}
-\frac{309181\zeta_2^2}{3240}
+\frac{533\zeta_2^2\zeta_3}{45}
-\frac{127\zeta_2^3}{3}
\biggr)\Biggr ] \nonumber \\ 
{\cal F}_2^g &=& {\cal F}_2^g|_{\frac{1}{\eps^4}} + \ldots +{\cal F}_2^g|_{\eps^3}\nonumber \\ 
&+& \CA^2 \Biggl [ \eps^4\biggl(+\frac{1371828689}{209952}
-264\zeta_{5,3}
+\frac{56155\zeta_7}{42}
-\frac{161266\zeta_5}{135}
-\frac{5108944\zeta_3}{729}
+\frac{1690\zeta_3\zeta_5}{3}
\nonumber \\ &&\hspace{1cm}
+\frac{85559\zeta_3^2}{81}
-\frac{219275\zeta_2}{1944}
-\frac{1001\zeta_2\zeta_5}{5}
+\frac{11858\zeta_2\zeta_3}{27}
-\frac{1547\zeta_2\zeta_3^2}{9}
-\frac{187733\zeta_2^2}{180}
\nonumber \\ &&\hspace{1cm}
+\frac{22781\zeta_2^2\zeta_3}{90}
+\frac{123079\zeta_2^3}{1260}
+\frac{50419\zeta_2^4}{100}
\biggr)\Biggr ] \nonumber \\ 
&+& \CA\NF \Biggl [ \eps^4\biggl(-\frac{232282297}{104976}
+\frac{229\zeta_7}{21}
-\frac{24518\zeta_5}{135}
-\frac{301886\zeta_3}{729}
+\frac{22060\zeta_3^2}{81}
\nonumber \\ &&\hspace{1cm}
+\frac{98791\zeta_2}{972}
+\frac{342\zeta_2\zeta_5}{5}
+\frac{2978\zeta_2\zeta_3}{27}
-\frac{40148\zeta_2^2}{405}
+\frac{517\zeta_2^2\zeta_3}{5}
+\frac{2167\zeta_2^3}{630}
\biggr)\Biggr ] \nonumber \\ 
&+& \CF\NF \Biggl [ \eps^4\biggl(-\frac{19296691}{7776}
-254\zeta_7
+\frac{22948\zeta_5}{45}
+\frac{192068\zeta_3}{81}
-460\zeta_3^2
+\frac{75305\zeta_2}{648}
\nonumber \\ &&\hspace{1cm}
-32\zeta_2\zeta_5
-\frac{5716\zeta_2\zeta_3}{27}
+\frac{585929\zeta_2^2}{1620}
-\frac{6724\zeta_2^2\zeta_3}{45}
-\frac{2024\zeta_2^3}{105}
\biggr)\Biggr ] 
\end{eqnarray}
}

The unrenormalised three-loop quark form factor ${\cal F}_3^q$ through to (and including) ${\cal O}(\eps^0)$ is given in eq.~(5.4) of Ref.~\cite{GGHIS}. The pole contributions of ${\cal F}_3^q$ are also given in eq.~(3.7) of ref.~\cite{MMV1} while the finite parts of the $N_F^2$, $C_AN_F$ and $C_FN_F$ contributions are given in eq.~(6) of ref.~\cite{MMV2}.  The 
finite  $N_{F,V}$ contribution 
could already be inferred from~\cite{MVVsplit}.
The remaining finite contributions are also given in eqs.~(8) and (9) of ref.~\cite{BCSSS}.
The ${\cal O}(\eps^1)$ and ${\cal O}(\eps^2)$ contributions are given by,

{\allowdisplaybreaks
\begin{eqnarray}
{\cal F}_3^q &=& {\cal F}_3^q|_{\frac{1}{\eps^6}} + \ldots +{\cal F}_3^q|_{\eps^0}\nonumber \\
+ \CF^3 \Biggl [ 
  +&\eps&\biggl(
-\frac{343393}{48}
-\frac{11896\zeta_{7}}{7}
+\frac{22349\zeta_{5}}{3}
+\frac{40835\zeta_{3}}{6}
-1203\zeta_{3}^2
-\frac{105553\zeta_{2}}{24}
\nonumber \\ &&\hspace{1cm}
-\frac{7858\zeta_{2}\zeta_{5}}{15}
+\frac{6083\zeta_{2}\zeta_{3}}{6}
+\frac{36693\zeta_{2}^2}{40}
-\frac{3931\zeta_{2}^2\zeta_{3}}{6}
+\frac{321227\zeta_{2}^3}{840}
\biggr) \nonumber \\  
+&\eps^2&\biggl(
-\frac{2512115}{96}
+\frac{4160\zeta_{5,3}}{3}
+\frac{45168\zeta_{7}}{7}
+\frac{716537\zeta_{5}}{15}
-\frac{137417\zeta_{3}}{12}
\nonumber \\ &&\hspace{1cm}
-\frac{33148\zeta_{3}\zeta_{5}}{3}
+\frac{12749\zeta_{3}^2}{6}
-\frac{797995\zeta_{2}}{48}
-\frac{12361\zeta_{2}\zeta_{5}}{5}
+\frac{18469\zeta_{2}\zeta_{3}}{2}
\nonumber \\ &&\hspace{1cm}
+1985\zeta_{2}\zeta_{3}^2
+\frac{7653\zeta_{2}^2}{80}
-\frac{15491\zeta_{2}^2\zeta_{3}}{20}
+\frac{1147979\zeta_{2}^3}{240}
-\frac{74208727\zeta_{2}^4}{50400}
\biggr)\Biggr ] \nonumber \\  
+ \CF^2 \CA \Biggl [ 
  +&\eps&\biggl(
+\frac{783459131}{34992}
-1349\zeta_{7}
-\frac{1894909\zeta_{5}}{270}
-\frac{1259477\zeta_{3}}{54}
+\frac{85649\zeta_{3}^2}{18}
\nonumber \\ &&\hspace{1cm}
+\frac{19394303\zeta_{2}}{1944}
+\frac{4851\zeta_{2}\zeta_{5}}{5}
-\frac{195175\zeta_{2}\zeta_{3}}{108}
-\frac{15062939\zeta_{2}^2}{6480}
\nonumber \\ &&\hspace{1cm}
+\frac{9751\zeta_{2}^2\zeta_{3}}{20}
-\frac{1811231\zeta_{2}^3}{15120}
\biggr) \nonumber \\  
+&\eps^2&\biggl(
+\frac{16308475427}{209952}
-\frac{15472\zeta_{5,3}}{15}
+\frac{415489\zeta_{7}}{42}
-\frac{7913725\zeta_{5}}{162}
-\frac{27356135\zeta_{3}}{324}
\nonumber \\ &&\hspace{1cm}
+\frac{72904\zeta_{3}\zeta_{5}}{15}
+\frac{2174933\zeta_{3}^2}{108}
+\frac{521534243\zeta_{2}}{11664}
+\frac{53128\zeta_{2}\zeta_{5}}{15}
-\frac{5620115\zeta_{2}\zeta_{3}}{324}
\nonumber \\ &&\hspace{1cm}
-1425\zeta_{2}\zeta_{3}^2
-\frac{161423233\zeta_{2}^2}{19440}
+\frac{1083953\zeta_{2}^2\zeta_{3}}{180}
-\frac{211343621\zeta_{2}^3}{90720}
\nonumber \\ &&\hspace{1cm}
-\frac{22796551\zeta_{2}^4}{63000}
\biggr)\Biggr ] \nonumber \\  
+ \CF \CA^2 \Biggl [ 
  +&\eps&\biggl(
-\frac{458292965}{26244}
-\frac{211\zeta_{7}}{18}
+\frac{15601\zeta_{5}}{5}
+\frac{42813461\zeta_{3}}{2916}
-\frac{71734\zeta_{3}^2}{27}
\nonumber \\ &&\hspace{1cm}
-\frac{52068575\zeta_{2}}{8748}
-\frac{1568\zeta_{2}\zeta_{5}}{9}
+\frac{13139\zeta_{2}\zeta_{3}}{27}
+\frac{4467743\zeta_{2}^2}{3240}
-\frac{4408\zeta_{2}^2\zeta_{3}}{45}
\nonumber \\ &&\hspace{1cm}
-\frac{8009\zeta_{2}^3}{945}
\biggr) \nonumber \\  
+&\eps^2&\biggl(
-\frac{34868838031}{472392}
-\frac{3592\zeta_{5,3}}{45}
-\frac{176495\zeta_{7}}{36}
+\frac{18727307\zeta_{5}}{810}
+\frac{405838949\zeta_{3}}{5832}
\nonumber \\ &&\hspace{1cm}
+\frac{568\zeta_{3}\zeta_{5}}{3}
-\frac{820579\zeta_{3}^2}{54}
-\frac{1546106255\zeta_{2}}{52488}
-\frac{23456\zeta_{2}\zeta_{5}}{15}
+\frac{2116327\zeta_{2}\zeta_{3}}{324}
\nonumber \\ &&\hspace{1cm}
+\frac{2896\zeta_{2}\zeta_{3}^2}{9}
+\frac{167549\zeta_{2}^2}{27}
-3805\zeta_{2}^2\zeta_{3}
+\frac{201469\zeta_{2}^3}{216}
+\frac{6341548\zeta_{2}^4}{23625}
\biggr)\Biggr ] \nonumber \\  
+ \CF^2\NF \Biggl [ 
  +&\eps&\biggl(
-\frac{50187205}{17496}
+\frac{5863\zeta_{5}}{135}
+\frac{929587\zeta_{3}}{243}
-\frac{5771\zeta_{3}^2}{9}
-\frac{1263505\zeta_{2}}{972}
\nonumber \\ &&\hspace{1cm}
-\frac{8515\zeta_{2}\zeta_{3}}{54}
+\frac{821749\zeta_{2}^2}{3240}
-\frac{875381\zeta_{2}^3}{7560}
\biggr) \nonumber \\  
+&\eps^2&\biggl(
-\frac{861740653}{104976}
-\frac{294430\zeta_{7}}{63}
+\frac{167299\zeta_{5}}{81}
+\frac{32307433\zeta_{3}}{1458}
-\frac{208487\zeta_{3}^2}{54}
\nonumber \\ &&\hspace{1cm}
-\frac{32868205\zeta_{2}}{5832}
+\frac{953\zeta_{2}\zeta_{5}}{15}
-\frac{152867\zeta_{2}\zeta_{3}}{162}
+\frac{17061119\zeta_{2}^2}{9720}
-\frac{172799\zeta_{2}^2\zeta_{3}}{180}
\nonumber \\ &&\hspace{1cm}
-\frac{4769039\zeta_{2}^3}{6480}
\biggr)\Biggr ] \nonumber \\  
+ \CF\CA\NF \Biggl [ 
  +&\eps&\biggl(
+\frac{24570881}{4374}
-\frac{28156\zeta_{5}}{45}
-\frac{2418896\zeta_{3}}{729}
+\frac{10816\zeta_{3}^2}{27}
+\frac{7137385\zeta_{2}}{4374}
\nonumber \\ &&\hspace{1cm}
+\frac{2674\zeta_{2}\zeta_{3}}{27}
-\frac{352559\zeta_{2}^2}{1620}
+\frac{17324\zeta_{2}^3}{945}
\biggr) \nonumber \\  
+&\eps^2&\biggl(
+\frac{5509319623}{236196}
+1170\zeta_{7}
-\frac{622178\zeta_{5}}{135}
-\frac{79031137\zeta_{3}}{4374}
+\frac{218296\zeta_{3}^2}{81}
\nonumber \\ &&\hspace{1cm}
+\frac{102669593\zeta_{2}}{13122}
+\frac{3272\zeta_{2}\zeta_{5}}{15}
+\frac{11939\zeta_{2}\zeta_{3}}{81}
-\frac{3829919\zeta_{2}^2}{3240}
+\frac{9572\zeta_{2}^2\zeta_{3}}{15}
\nonumber \\ &&\hspace{1cm}
+\frac{74461\zeta_{2}^3}{5670}
\biggr)\Biggr ] \nonumber \\  
+ \CF\NF^2 \Biggl [ 
  +&\eps&\biggl(
-\frac{2913928}{6561}
+\frac{2248\zeta_{5}}{135}
+\frac{2108\zeta_{3}}{27}
-\frac{24950\zeta_{2}}{243}
+\frac{68\zeta_{2}\zeta_{3}}{9}
-\frac{3901\zeta_{2}^2}{810}
\biggr) \nonumber \\  
+&\eps^2&\biggl(
-\frac{109448624}{59049}
+\frac{52828\zeta_{5}}{405}
+\frac{848300\zeta_{3}}{2187}
-\frac{1156\zeta_{3}^2}{81}
-\frac{338858\zeta_{2}}{729}
\nonumber \\ &&\hspace{1cm}
+\frac{1598\zeta_{2}\zeta_{3}}{27}
-\frac{2573\zeta_{2}^2}{90}
+\frac{44651\zeta_{2}^3}{5670}
\biggr)\Biggr ] \nonumber \\  
\lefteqn{+ \CF\NFV\NN \times }\nonumber \\ 
\Biggl [ 
  +&\eps&\biggl(
+\frac{170}{3}
+\frac{752\zeta_{5}}{9}
+\frac{94\zeta_{3}}{9}
-\frac{344\zeta_{3}^2}{3}
+\frac{260\zeta_{2}}{3}
+30\zeta_{2}\zeta_{3}
-\frac{196\zeta_{2}^2}{15}
-\frac{9728\zeta_{2}^3}{315}
\biggr) \nonumber \\  
+&\eps^2&\biggl(
+\frac{1460}{3}
-\frac{4271\zeta_{7}}{3}
+\frac{12970\zeta_{5}}{27}
+\frac{2501\zeta_{3}}{27}
-\frac{748\zeta_{3}^2}{9}
+\frac{4345\zeta_{2}}{9}
\nonumber \\ &&\hspace{1cm}
-\frac{256\zeta_{2}\zeta_{5}}{3}
+\frac{239\zeta_{2}\zeta_{3}}{3}
-\frac{3677\zeta_{2}^2}{45}
-\frac{392\zeta_{2}^2\zeta_{3}}{3}
+\frac{85244\zeta_{2}^3}{945}
\biggr)\Biggr ]  
\end{eqnarray}
}
Note that last colour factor 
is generated by graphs where the virtual gauge boson does 
not couple directly to the final-state quarks.  This 
contribution is denoted by $N_{F,V}$ and is proportional to the charge 
weighted sum of the quark flavours.  In the case of purely 
electromagnetic interactions, we find,
\begin{equation}
N_{F,\gamma} = \frac{\sum_q e_q}{e_q}. 
\end{equation}

The unrenormalised three-loop gluon form factor through to (and including) ${\cal O}(\eps^0)$ is given in eq. (5.5) of Ref.~\cite{GGHIS}.  The divergent parts are also given in eq.~(8) of ref.~\cite{MMV2} while the finite contributions are given in eq.~(10) of ref.~\cite{BCSSS}.
The ${\cal O}(\eps^1)$ and ${\cal O}(\eps^2)$ contributions for ${\cal F}_3^g$ are given by,
{\allowdisplaybreaks
\begin{eqnarray}
{\cal F}_3^g &=& {\cal F}_3^g|_{\frac{1}{\eps^6}} + \ldots + {\cal F}_3^g|_{\eps^0}\nonumber \\
+ \CA^3 \Biggl [ 
  +&\eps&\biggl(
+\frac{270573319}{26244}
-\frac{385579\zeta_{7}}{126}
+\frac{389159\zeta_{5}}{135}
-\frac{3601570\zeta_{3}}{729}
+\frac{74899\zeta_{3}^2}{54}
\nonumber \\ &&\hspace{1cm}
-\frac{446863\zeta_{2}}{4374}
+\frac{2449\zeta_{2}\zeta_{5}}{9}
-\frac{34093\zeta_{2}\zeta_{3}}{54}
-\frac{40819\zeta_{2}^2}{180}
-\frac{47803\zeta_{2}^2\zeta_{3}}{180}
\nonumber \\ &&\hspace{1cm}
+\frac{7200127\zeta_{2}^3}{15120}
\biggr) \nonumber \\  
+&\eps^2&\biggl(
+\frac{30151577675}{472392}
+\frac{12392\zeta_{5,3}}{45}
+\frac{2169431\zeta_{7}}{126}
+\frac{3101341\zeta_{5}}{405}
-\frac{59902487\zeta_{3}}{1458}
\nonumber \\ &&\hspace{1cm}
-\frac{89996\zeta_{3}\zeta_{5}}{15}
+\frac{16453\zeta_{3}^2}{2}
-\frac{108299125\zeta_{2}}{26244}
-\frac{6897\zeta_{2}\zeta_{5}}{10}
-\frac{80255\zeta_{2}\zeta_{3}}{27}
\nonumber \\ &&\hspace{1cm}
+\frac{7936\zeta_{2}\zeta_{3}^2}{9}
-\frac{34875497\zeta_{2}^2}{9720}
+\frac{714109\zeta_{2}^2\zeta_{3}}{360}
+\frac{12226469\zeta_{2}^3}{5040}
-\frac{1183759981\zeta_{2}^4}{756000}
\biggr)\Biggr ] \nonumber \\  
+ \CA^2\NF \Biggl [ 
  +&\eps&\biggl(
-\frac{48658741}{8748}
-\frac{10066\zeta_{5}}{45}
+\frac{349918\zeta_{3}}{729}
-\frac{11657\zeta_{3}^2}{27}
+\frac{904045\zeta_{2}}{4374}
\nonumber \\ &&\hspace{1cm}
+\frac{791\zeta_{2}\zeta_{3}}{9}
-\frac{34931\zeta_{2}^2}{1620}
-\frac{52283\zeta_{2}^3}{1080}
\biggr) \nonumber \\  
+&\eps^2&\biggl(
-\frac{15039308929}{472392}
-\frac{14271\zeta_{7}}{7}
-\frac{391564\zeta_{5}}{405}
+\frac{13422322\zeta_{3}}{2187}
-\frac{76349\zeta_{3}^2}{81}
\nonumber \\ &&\hspace{1cm}
+\frac{66386911\zeta_{2}}{26244}
+\frac{307\zeta_{2}\zeta_{5}}{5}
+\frac{31849\zeta_{2}\zeta_{3}}{81}
+\frac{373234\zeta_{2}^2}{1215}
-\frac{104327\zeta_{2}^2\zeta_{3}}{180}
\nonumber \\ &&\hspace{1cm}
-\frac{6878021\zeta_{2}^3}{22680}
\biggr)\Biggr ] \nonumber \\  
+ \CA\CF\NF \Biggl [ 
  +&\eps&\biggl(
-\frac{10508593}{2916}
+\frac{17092\zeta_{5}}{27}
+\frac{240934\zeta_{3}}{243}
+\frac{4064\zeta_{3}^2}{9}
+\frac{8869\zeta_{2}}{54}
\nonumber \\ &&\hspace{1cm}
+\frac{640\zeta_{2}\zeta_{3}}{9}
+\frac{28823\zeta_{2}^2}{270}
+\frac{23624\zeta_{2}^3}{315}
\biggr) \nonumber \\  
+&\eps^2&\biggl(
-\frac{418631245}{17496}
+\frac{16658\zeta_{7}}{9}
+\frac{386102\zeta_{5}}{81}
+\frac{4492979\zeta_{3}}{729}
+\frac{17176\zeta_{3}^2}{27}
\nonumber \\ &&\hspace{1cm}
+\frac{163523\zeta_{2}}{108}
-496\zeta_{2}\zeta_{5}
+\frac{3500\zeta_{2}\zeta_{3}}{9}
+\frac{437599\zeta_{2}^2}{540}
+\frac{3148\zeta_{2}^2\zeta_{3}}{5}
\nonumber \\ &&\hspace{1cm}
+\frac{157424\zeta_{2}^3}{315}
\biggr)\Biggr ] \nonumber \\  
+ \CF^2\NF \Biggl [ 
  +&\eps&\biggl(
+\frac{18613}{54}
-\frac{3080\zeta_{5}}{3}
+\frac{10552\zeta_{3}}{9}
-272\zeta_{3}^2
-\frac{74\zeta_{2}}{3}
-16\zeta_{2}\zeta_{3}
+\frac{328\zeta_{2}^2}{5}
-\frac{35648\zeta_{2}^3}{315}
\biggr) \nonumber \\  
+&\eps^2&\biggl(
+\frac{383765}{162}
-\frac{8828\zeta_{7}}{3}
-\frac{35956\zeta_{5}}{9}
+\frac{229772\zeta_{3}}{27}
-\frac{6400\zeta_{3}^2}{3}
-\frac{4109\zeta_{2}}{18}
\nonumber \\ &&\hspace{1cm}
+560\zeta_{2}\zeta_{5}
-276\zeta_{2}\zeta_{3}
+764\zeta_{2}^2
-\frac{1232\zeta_{2}^2\zeta_{3}}{3}
-\frac{796168\zeta_{2}^3}{945}
\biggr)\Biggr ] \nonumber \\  
+ \CA\NF^2 \Biggl [ 
  +&\eps&\biggl(
+\frac{16823771}{26244}
+\frac{9368\zeta_{5}}{135}
+\frac{5440\zeta_{3}}{27}
-\frac{30283\zeta_{2}}{1458}
-\frac{988\zeta_{2}\zeta_{3}}{27}
+\frac{14018\zeta_{2}^2}{405}
\biggr) \nonumber \\  
+&\eps^2&\biggl(
+\frac{1534229129}{472392}
+\frac{33136\zeta_{5}}{81}
+\frac{1698929\zeta_{3}}{2187}
-\frac{17908\zeta_{3}^2}{81}
-\frac{1822421\zeta_{2}}{8748}
\nonumber \\ &&\hspace{1cm}
-\frac{15928\zeta_{2}\zeta_{3}}{81}
+\frac{20009\zeta_{2}^2}{135}
+\frac{12851\zeta_{2}^3}{5670}
\biggr)\Biggr ] \nonumber \\  
+ \CF\NF^2 \Biggl [ 
  +&\eps&\biggl(
+\frac{196900}{243}
-\frac{800\zeta_{5}}{9}
-\frac{4208\zeta_{3}}{9}
-54\zeta_{2}
+\frac{112\zeta_{2}\zeta_{3}}{3}
-\frac{2464\zeta_{2}^2}{45}
\biggr) \nonumber \\  
+&\eps^2&\biggl(
+\frac{6322579}{1458}
-\frac{17600\zeta_{5}}{27}
-\frac{223756\zeta_{3}}{81}
+\frac{3232\zeta_{3}^2}{9}
-\frac{9626\zeta_{2}}{27}
+\frac{2464\zeta_{2}\zeta_{3}}{9}
\nonumber \\ &&\hspace{1cm}
-\frac{4913\zeta_{2}^2}{15}
+\frac{248\zeta_{2}^3}{63}
\biggr)\Biggr ]
\end{eqnarray}
}

The renormalised form factors are directly related to the unrenormalised form factors and details on how to extract the renormalised form factors to this order are given in section 2 of Ref.~\cite{GGHIS}.

In this letter, we computed the three-loop quark and gluon form factors through to {\cal O}$(\eps^2)$ in the
dimensional regularisation parameter. These contributions are relevant in the study of the infrared singularity structure at four loops. In particular, the ${\cal O}(\e)$ terms of 
the three-loop form factors are required for the extraction of the four-loop 
quark and gluon collinear anomalous dimensions.  The ${\cal O}(\e^2)$ terms contribute 
to the finite part of the infrared-subtraction of the form factors at four loops. It is this 
infrared-subtracted finite part which is relevant for the study of the next-to-next-to-next-to-next-to-leading (N$^4$LO) Drell-Yan and Higgs production processes.   In particular, the ${\cal O}(\e^2)$ three-loop contributions represent a finite ingredient to these processes at four-loops.

\section*{Acknowledgements}
We thank Daniel Ma\^{\i}tre and Volodya Smirnov for useful comments regarding the conversion of $\zeta_{-6,-2}$ to $\zeta_{5,3}$. This research was supported in part by
the Swiss National Science Foundation (SNF) under contract 200020-126691, by the 
Forschungskredit der Universit\"at Z\"urich,  the UK Science and Technology Facilities
Council, by the European Commission's Marie-Curie Research Training Network
under contract MRTN-CT-2006-035505 `Tools and Precision Calculations for Physics
Discoveries at Colliders', by the Helmholtz Alliance ``Physics at the Terascale'',
and by Deutsche Forschungsgemeinschaft (DFG SCHR 993/2-1).
EWNG gratefully acknowledges the support of the
Wolfson Foundation and the Royal Society.

\end{document}